\documentclass{vietnam}
\usepackage{amsmath}
\usepackage{amssymb}
\usepackage[dvipsnames]{xcolor}

\def\Journal#1#2#3#4{{#1} {\bf #2}, #3 (#4)}

\makeatletter
\newcommand\notsotiny{\@setfontsize\notsotiny\@viipt\@viiipt}
\makeatother

\def\NPB{{\em Nucl.\ Phys.} B}
\def\PLB{{\em Phys.\ Lett.}  B}
\def\PR{\em Phys.\ Rev.}
\def\PRL{\em Phys.\ Rev.\ Lett.}
\def\PRD{{\em Phys.\ Rev.} D}

\def\PRSA{{\em Proc.\ R. Soc.} A}
\def\PhysRept{{\em Phys.\ Rept.}}

\def\be{\begin{equation}}
\def\ee{\end{equation}}
\def\bea{\begin{eqnarray}}
\def\eea{\end{eqnarray}}

\newcommand{\Photo}{\includegraphics[height=35mm]{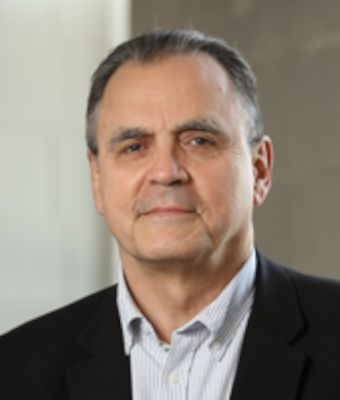}  }

\begin{document}
\vspace*{4cm}
\title{Muon magnetic anomaly: experimental status and prospects}

\author{ Dinko Po\v{c}ani\'c }

\address{Department of Physics, University of Virginia,
         Charlottesvile, VA 22904-4714, USA}

\maketitle\abstracts{
  The past five years have brought significant new results on the muon
  magnetic anomaly, $a_{\mu} = (g_{\mu}-2)/2$, and on the hadronic vacuum
  polarization (HVP) contribution dominating the uncertainty $\Delta a_{\mu}$.
  Serious tension has emerged between the experimental and standard model (SM)
  values for $a_{\mu}$, as well as between the SM and the first precise
  lattice QCD values.  We review the current experimental and theoretical
  status of $a_{\mu}$, along with the prospects for new results, focusing on
  MUonE, a new experiment at CERN, aiming to evaluate the leading order
  contribution to $a_{\mu}^{\text{HVP}}$ in a direct measurement of muonic
  Bhabha scattering.
                    }

\section{The physics of the muon magnetic anomaly \label{sec:motiv}}

The magnetic moments of the electron and muon, given by their gyromagnetic
ratios $g_{e,\mu}$,
\begin{equation*}
  \vec{\mu}_{\ell} = g_{\ell} \left(\frac{q}{2m_{\ell}}\right) \vec{S}\,,
  \quad \text{where} \quad  g_{\ell} = 2(1+a_{\ell})\,, \quad \text{and}\quad
  \ell = e,\mu\,,
  \label{eq:MDM}
\end{equation*}
have played an important role in the development of the standard model (SM).
The Dirac theory\,\cite{Dirac:1928ej} predicted that $g_e =2$, resolving a
decades-long search for a fundamental explanation of this anomaly.  About 20
years later, Schwinger\,\cite{Schwinger:1948iu} proposed an additional
contribution to $\mu_e$ from a radiative correction, predicting the
anomaly\,\footnote{The quantity $a_{\ell}$ is the magnetic anomaly, but is
also often referred to simply as the ``anomaly'' or as the ``anomalous
magnetic moment''.}  $a_e = \alpha/2\pi \simeq 0.00116$ in good agreement with
contemporary experiments.\cite{Kusch:1948aa} A decade later, a precise
experiment\,\cite{Garwin:1960zz} confirmed Schwinger's prediction for the muon
anomaly, and thereby established for the first time that, in a magnetic field,
a muon behaves simply like a massive electron, which, in turn, paved the way
for the notion of lepton generations.

\begin{figure}[t]
  \begin{center}
    \parbox{0.50\linewidth}{
      \includegraphics[width=\linewidth]{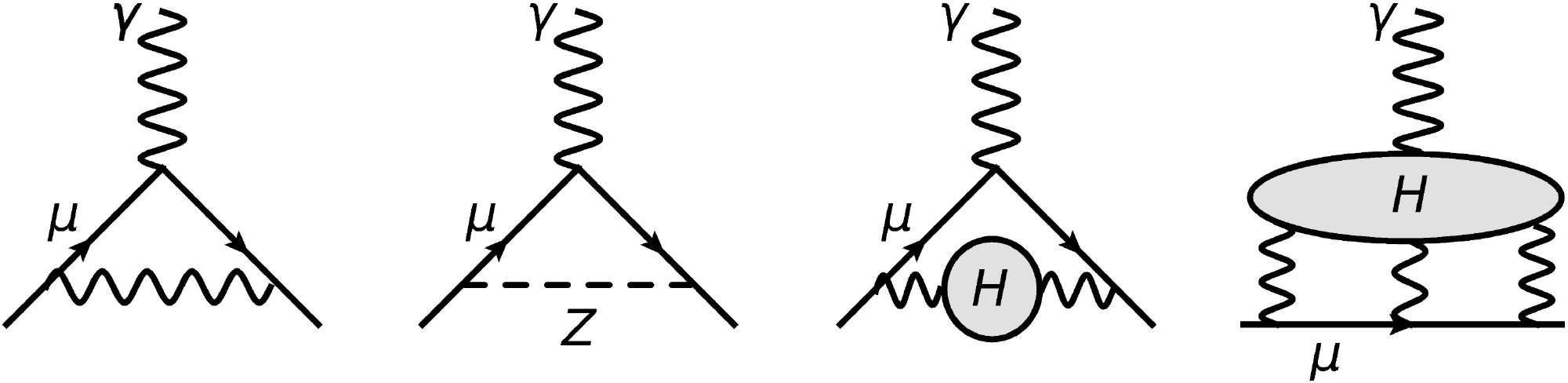}
                           }
    \caption{Feynman diagrams of representative SM contributions to $a_{\mu}$.
      From left to right: first-order QED and weak processes, leading-order
      hadronic (H) vacuum polarization and hadronic light-by-light
      contributions.}\label{fig:feyn-main}
  \end{center}
\end{figure}

In the SM, the muon anomaly receives contributions from electromagnetic,
strong and weak interactions that arise from virtual effects involving
photons, leptons, hadrons, as well as the $W$, $Z$, and Higgs
bosons,\cite{Jegerlehner:2017gek} as shown in Fig.~\ref{fig:feyn-main}.
While $a_e$, the electron magnetic anomaly, is experimentally determined far
more precisely than $a_{\mu}$, the latter is $(m_\mu/m_e)^2 \simeq$\ 43,000
times more sensitive to contributions from heavy additions to the SM through
processes similar to those illustrated in Fig.~\ref{fig:feyn-main}.  This
extraordinary sensitivity has motivated the international theory community to
embark on a comprehensive program of evaluating $a_{\mu}$ that culminated in
the 2020 prediction of $a_\mu(\text{SM}) = 116\,591\,810(43) \times 10^{-11}$
(0.37\,ppm), based on calculations of previously unparalleled precision,
published in the 2020 Muon $g-2$ Theory Initiative White
Paper (TIWP).\cite{Aoyama:2020ynm} The theoretical uncertainty is dominated by the
hadronic contributions, particularly by the hadronic vacuum polarization (HVP)
term, followed by the hadronic light by light (HLbL).  The theoretical
uncertainty is dominated by the contribution from the leading order HVP term,
as summarized in Table~\ref{tab:amu_SM-theo}.  The key ingredient in the SM
theoretical prediction for $a_\mu$ is the critically compiled comprehensive
data set on $\sigma(e^+ + e^- \to \text{hadrons}) = \sigma_{\text{had}}(s)$
cross sections, that are subjected to a dispersion-relation integral
transformation (based on the optical theorem and analiticity) to arrive at the
HVP contribution $a_\mu^{\text{HVP}}$ (see ref.~\cite{Aoyama:2020ynm} for
details).  We note that the uncertainties assigned to the QED and weak
interaction contributions are insignificant in comparison to those of the
hadronic terms in $a_\mu^{\text{SM}}$.
\begin{table}[b]
  \caption[]{Top-level summary of the contributions to the SM value of $a_\mu$
    from each of the three fundamental interactions: electromagnetic (QED),
    weak, and hadronic (HVP and HLbL), along with their respective
    uncertainties, from the 2020 Muon $g-2$ Theory Initiative White Paper
    \cite{Aoyama:2020ynm} (further details are provided in the reference).}
  \label{tab:amu_SM-theo}
  \vspace*{4mm}
  \begin{center}
  \begin{tabular}{|l|r|r|}
      \hline \\[-12pt]
        $a_\mu$ term &
          { value \scriptsize $(\times \, 10^{-11})$ } & 
          uncertainty \\
      \hline
        QED    & 116,584,718.931              &  0.104 \\
        Weak  &         153.6\hspace*{10pt}  & 1.0\hspace*{10pt} \\
        HVP    &       6,845\hspace*{18pt}    & 40\hspace*{18pt} \\
        HLbL   &           92\hspace*{18pt}   & 18\hspace*{18pt} \\ 
      \hline
        Total SM  &  116,591,810\hspace*{18pt}  & 43\hspace*{18pt} \\
      \hline
  \end{tabular}
  \end{center} 
\end{table}

Meanwhile, the experimental evaluations of $a_\mu$ have been steadily
progressing toward ever more precise results.  The third and final muon $g-2$
experiment,\cite{CERN-Mainz-Daresbury:1978ccd} in a series of measurements at
CERN, achieved a precision of 10\,ppm for both $a_{\mu^+}$ and $a_{\mu^-}$, in
good mutual agreement.  CPT symmetry was assumed, and the results were
combined to give a 7.3\,ppm measurement, in agreement with theory.  This
result presented the first confirmation of the predicted 60 ppm contribution
to $a_\mu$ from hadronic vacuum polarization.  CERN-III experiment [5] used a
uniform-field storage ring (SR) and electric quadrupoles to provide vertical
contain- ment for the muons having the ``magic'' momentum of 3.1\,GeV/$c$.  At
this momentum, the muon spin precession is not affected by the electric field
from the focusing quadrupoles.

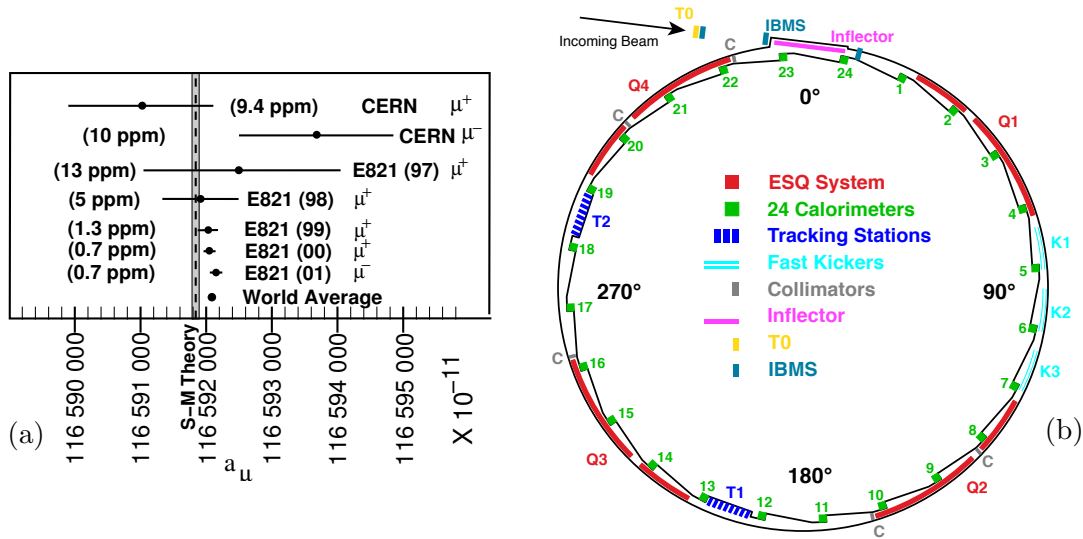
\begin{figure}[htb]
  \centering
    \parbox{0.4\linewidth}{
      \includegraphics[width=\linewidth]{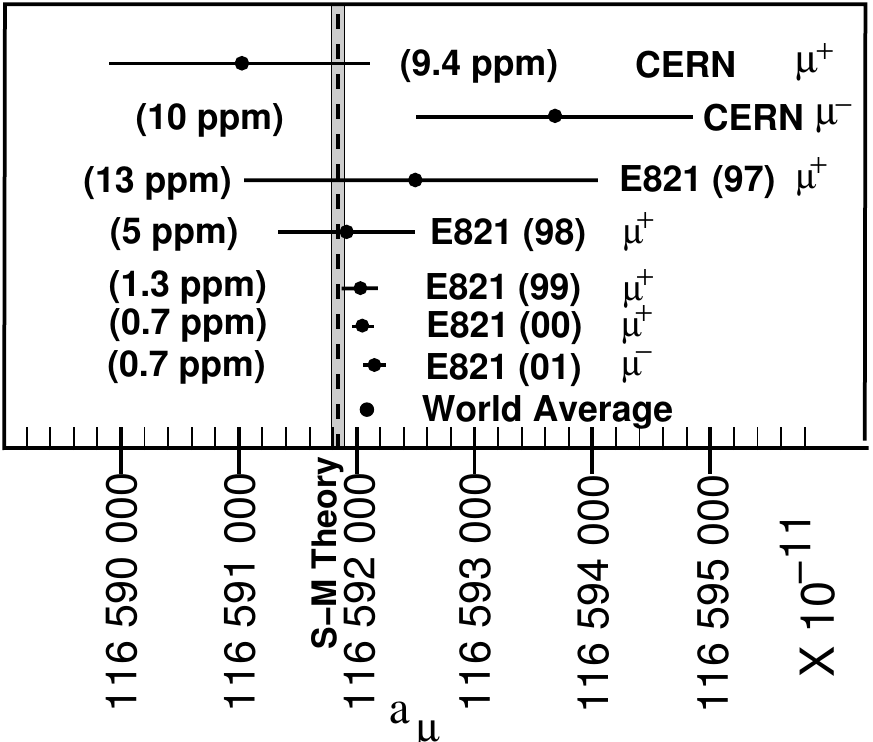} \\[-18pt]
      \begin{picture}(0,0)
        \put(0,10){(a)}
      \end{picture}
                          }
    \hspace*{9pt}
    \parbox{0.45\linewidth}{
      \includegraphics[width=\linewidth]{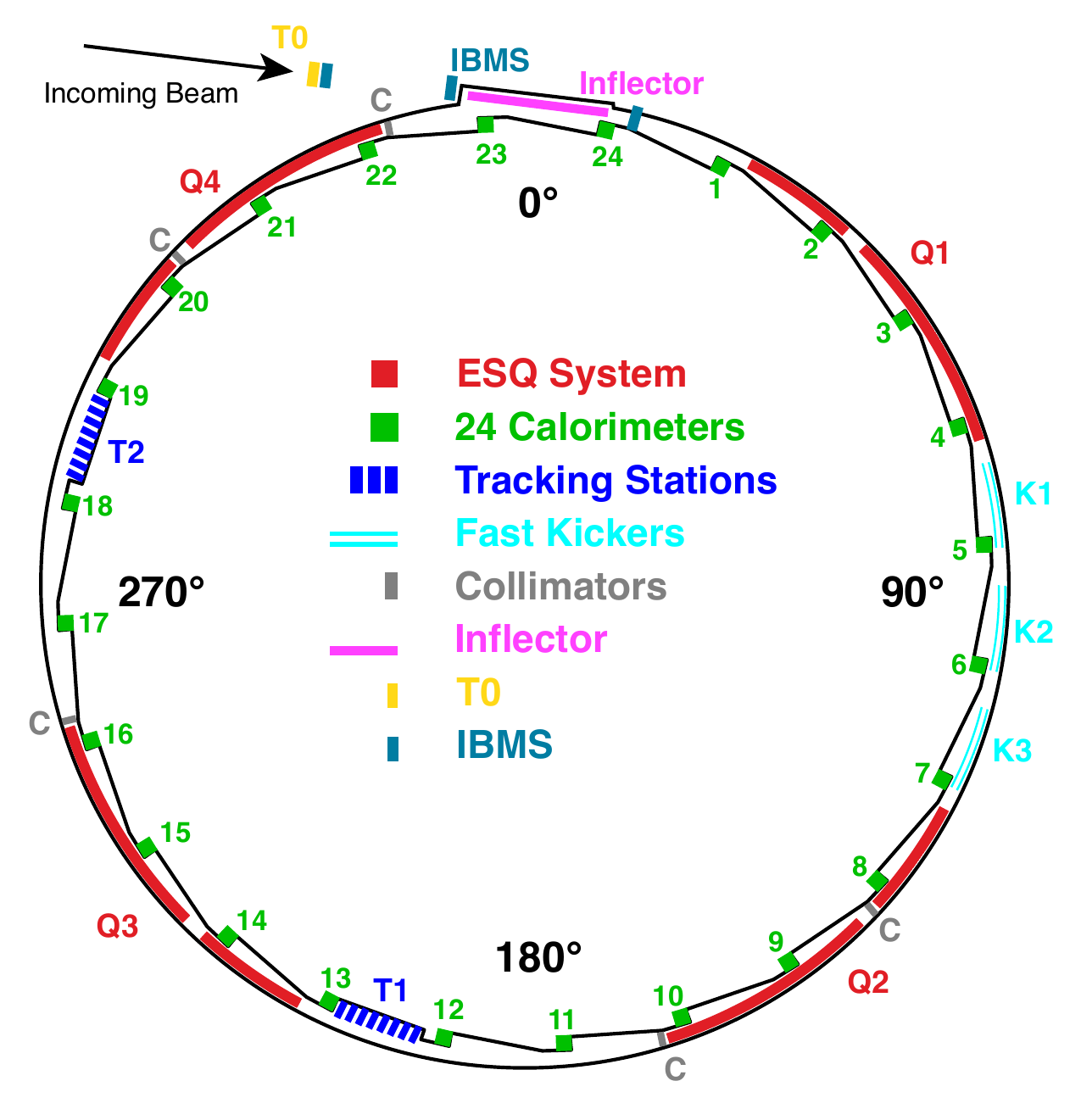} \\[-18pt]
      \begin{picture}(0,0)
        \put(190,40){(b)}
      \end{picture}
                           }
  \caption{(a) Evolution of the measured values for $a_\mu$ in experiments at
    CERN and BNL, compared to the SM prediction, as of 2006, the time of the
    final BNL E821 results.  (b) Layout of the Fermilab E989 storage ring that
    shares its design and most components with BNL E821, though with critical
    additions and upgrades.  ESQ denotes the electrostatic quadrupole beam
    focusing system comprising quadrupoles Q1--Q4, IBMS is the Injector beam
    monitoring system comprising scintillating fiber detectors, T0 is a thin
    plastic scintillation detector providing the time profile of arriving muon
    bunches, K1--K3 are three pulsed kicker plates needed to place the muon
    into a stable stored orbit, T1 and T2 are in-vacuo straw-tube tracking
    detectors, and, outside the vacuum, electromagnetic calorimeter stations,
    numbered 1--24, to detect the decay positrons.}
  \label{fig:g-2_comp_2006}
\end{figure}

The experimental effort was continued at the Brookhaven National Lab (BNL) in
USA, refining the CERN-III approach and introducing significant innovations.
In a series of measurements, the BNL E821 experiment improved the CERN-III
precision by over an order of magnitude, to 0,7\,ppm for both $a_{\mu^+}$ and
$a_{\mu^-}$.\cite{Muong-2:2006rrc} It is at this point, historically, that
appreciable differences emerged between the experimental and SM theoretical
evaluations of $a_\mu$, as seen in Figure~\ref{fig:g-2_comp_2006}(a).  The
discrepancy, $\sim 3\sigma$, was compelling in motivating the formation of an
expanded collaboration to perform a new round of measurements, improving
further on the CERN-BNL technique, this time at Fermilab (experiment FNAL
E989), where a more intense muon beam makes the desired improvement in
precision feasible in principle.

\section{\boldmath The Fermilab E989 Muon $g-2$ experiment \label{sec:E989}} 
The goal of the Fermilab E989 Muon $g-2$ experiment is to achieve a four-fold
improvement in precision of $a_\mu$ compared to the BNL E821 final result,
i.e., an overall uncertainty in $a_\mu$ of 140\,ppb, balanced between the
statistical and systematic uncertainties at 100\,ppb each.  The realization of
this goal has required that the collaboration upgrade an calibrate the E821
apparatus to a new, higher standard, and that a new muon beam line be designed
and constructed at FNAL.

Major components of the BNL E821 apparatus were transported from BNL on Long
Island, New York, to Fermilab, in Batavia, Illinois, most critically the bulky
$\sim 15$\,m diameter superconducting muon storage ring, in a storied trip
that involved helicopters, a barge voyage along the US east coast, through the
Gulf of Mexico, up the Mississippi and tributaries, as well as special trucks.
The apparatus was assembled and installed in the new MC-1 Hall of the Fermilab
Muon Campus.  Painstaking shimming and calibration of the storage ring
magnetic field was carried out in 2015-16 bringing the field uniformity to
within 50\,ppm, a stricter tolerance than was achieved in E821.  In parallel,
new apparatus, such as the 24 electromagnetic shower calorimeters, new
in-vacuo straw-tube tracking stations, a complement of beam detectors, etc.,
were prepared for the start of measurements.  By 2018, the collaboration
started taking first physics data, in Run-1.  In all, 6 runs were completed by
the end of physics data acquisition, all with $\mu^+$ beam.  To date, data
from Runs 1, 2 and 3 have been analyzed, and the results
published.\cite{Muong-2:2021ojo,Muong-2:2023cdq,Muong-2:2024hpx} The analysis of Runs 4-6 is
    ongoing, and its conclusion is anticipated soon.

It not feasible, given the format limitations, to provide in this space a
detailed discussion of all critical aspects of the experimental method,
calibrations, Monte Carlo simulations and analysis, that would reflect the
rigor required to reach the reported precision.  We, instead, point the reader
to Ref.~\cite{Muong-2:2024hpx} and references cited therein, and focus on the
basic principles of the measurement.

The complete breaking of chiral symmetry in the weak interaction forces the
spin vector of a positron with maximum energy (at the very endpoint of the
Michel spectrum) in the decay $\mu^+ \to \bar{\nu}_{\mu} + e^+ +\nu_e$, to be
aligned with $\vec{S}_\mu$, the spin of the decaying $\mu^+$.  If
$\vec{S}_\mu$ is aligned with $\vec{p}_\mu$, the muon momentum (3.1\,GeV/$c$),
the emitted $e^+$ is highly boosted, by a Lorentz $\gamma_e$ of 29.3.  If
$\vec{S}_\mu$ and $\vec{p}_\mu$ are anti-aligned, or misaligned, $\gamma_e$
drops precipitously.  Thus, $E_e$ provides an excellent measure of the
(mis)alignment of $\vec{S}_\mu$ and $\vec{p}_\mu$.  Classical electrodynamics
shows that the difference between $\omega_S$, the spin precession frequency in
an external magnetic field $B$, and $\omega_C$, cyclotron (momentum
precession) frequency, is determined by $a_\mu$ and $B$
\begin{equation}
  \vec{\omega}_a =\vec{\omega}_S -\vec{\omega}_C
    = -\frac{e}{m}\left[a_\mu\vec{B} -\left(a_\mu-\frac{1}{\gamma^2-1}\right)
      \frac{\vec{\beta} \times \vec{E}}{c}\right] \quad
    \xrightarrow[\gamma\to 29.3]{} \quad -\frac{e}{m}a_\mu\vec{B} \,,
    \label{eq:omega_a}
\end{equation}
i.e., the anomalous muon precession frequency becomes independent of the
electric field $\vec{E}$ at the ``magic'' momentum of $\gamma = 29.3$.  Under
these special circumstances, $a_\mu$ can be determined from two measurements,
of $\omega_a$ and $B$, where the latter can be expressed in terms of
$\omega_p$, the precession frequency of a proton in $\vec{B}$.  The precision
goal of E989, however, leads to the requirement that a number of corrections
be applied, not the least of which being that the measurement of $\vec{B}$ be
weighted by the muon beam intensity for all points of $(x,y)$ sampled by the
beam along the circumference of the muon SR.

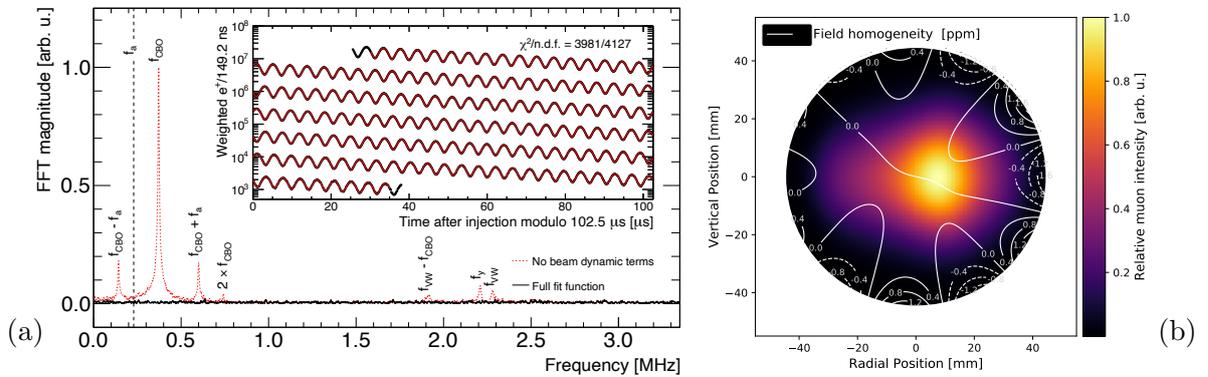
\begin{figure}[htb]
  \newlength{\plotheight}
  \setlength{\plotheight}{0.2\textheight}
  \begin{center}
    \includegraphics[height=\plotheight]{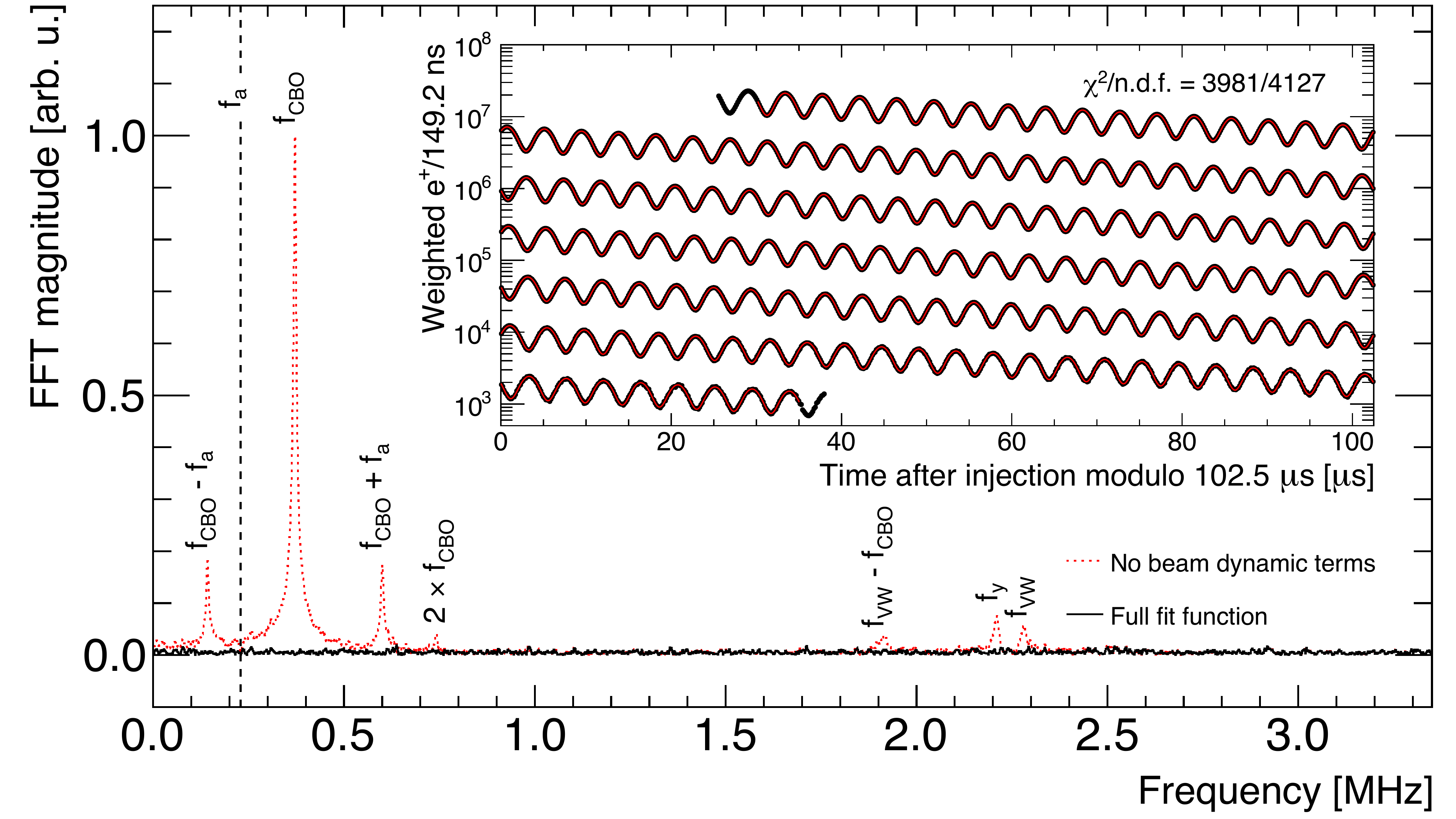} 
    \includegraphics[height=\plotheight]{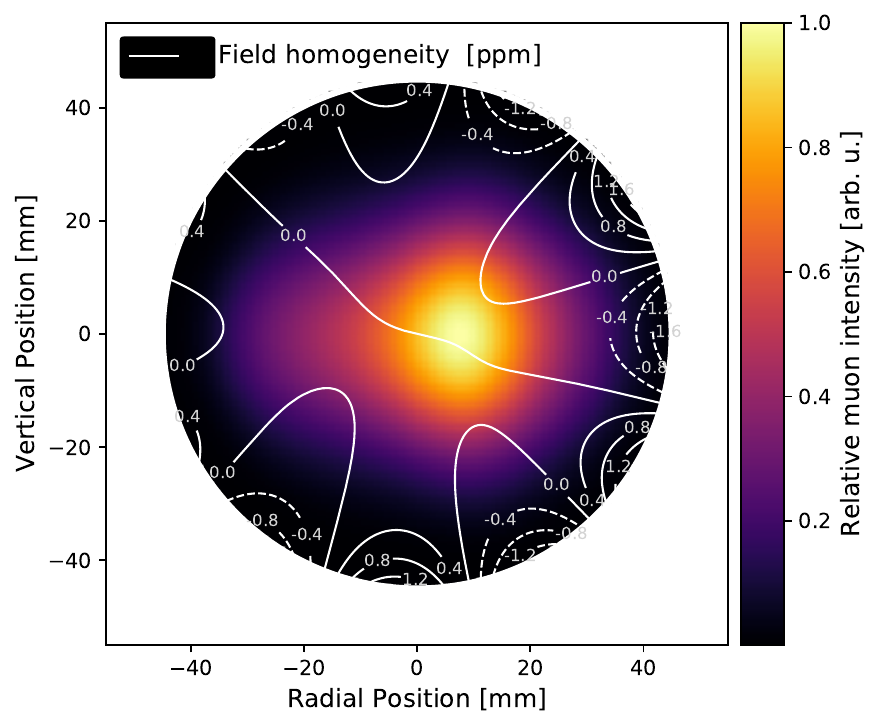} \\[-18pt]
    \begin{picture}(0,0)
      \put(-218,10){(a)}
      \put(+213,10){(b)}
    \end{picture}
    \caption{(a) Fourier transform of the residuals from a basic fit
      not including three beam dynamics parameters (red dashed line), and from
      the full fit (black line).  The peaks correspond to the missing betatron
      frequencies and muon losses.  Data are from the Run-3a data set.  Inset:
      corresponding asymmetry-weighted $e^+$ time spectrum (black line) with
      the full fit function (red line) overlaid.
      (b) Azimuthally averaged magnetic field contours overlaid on the time-
      and azimuthally averaged muon distribution for the Run-3b data set.
      For more details see  Refs.\protect\cite{Muong-2:2023cdq,Muong-2:2024hpx}
    \label{fig:wiggle_wtd-B_23}
            }
  \end{center}
\end{figure}

The principal experimental inputs into determining $a_\mu$ are illustrated in
Figure~\ref{fig:wiggle_wtd-B_23} based on data subsets collected in Run-2 and
Run-3.  Figure~\ref{fig:wiggle_wtd-B_23}(a) shows the ``wiggle'' plot of
measured $\omega_a$ oscillations (inset) and the accompanying Fourier
analysis.  Residuals from a basic fit not including all relevant beam dynamics
terms show pronounced peaks associated with the frequencies of the omitted
processes, primarily various modes of betatron oscillations.
Figure~\ref{fig:wiggle_wtd-B_23}(b) depicts an azimuthally averaged map of
magnetic field strength, weighted by the reconstructed muon beam particle
distribution in the radial-vertical displacement plane.  In spite of the
impressive field homogeneity, well under 1\,ppm level where the muons pass,
the beam density weighting remains necessary, on top of continuous field
monitoring and frequent periodic probe cross-calibrations.

The SR magnetic field is mapped every few days using a trolley instrumented
with nuclear magnetic resonance (NMR) probes housing petroleum jelly.  These
probes are calibrated using a retractable water-based cylindrical probe. This
enables the expression of the magnetic field in terms of the precession
frequency of shielded protons in a spherical sample $\omega'_p$, for which the
relation between precession frequency and magnetic field is precisely known.
Changes in the field between trolley measurements are tracked using 378 fixed
NMR probes embedded in the vacuum chamber walls above and below the muon
storage volume.  Dedicated instrumentation was in place to monitor transient
effects induced by the pulsing of kickers and electrostatic quadrupoles.  
It is particularly worth noting that all aspects of the data analysis were
carried out in parallel by several groups independently.  For example, the
$\omega_a$ ``wiggle'' plot analysis for Run-2 and Run-3 data was done by seven
groups using independent analysis routines.  Further, all E989 analysis is
doubly blinded, with unblinding taking place in controlled phases.

\begin{figure}[htb]
    \parbox{0.45\linewidth}{
      \includegraphics[width=\linewidth]{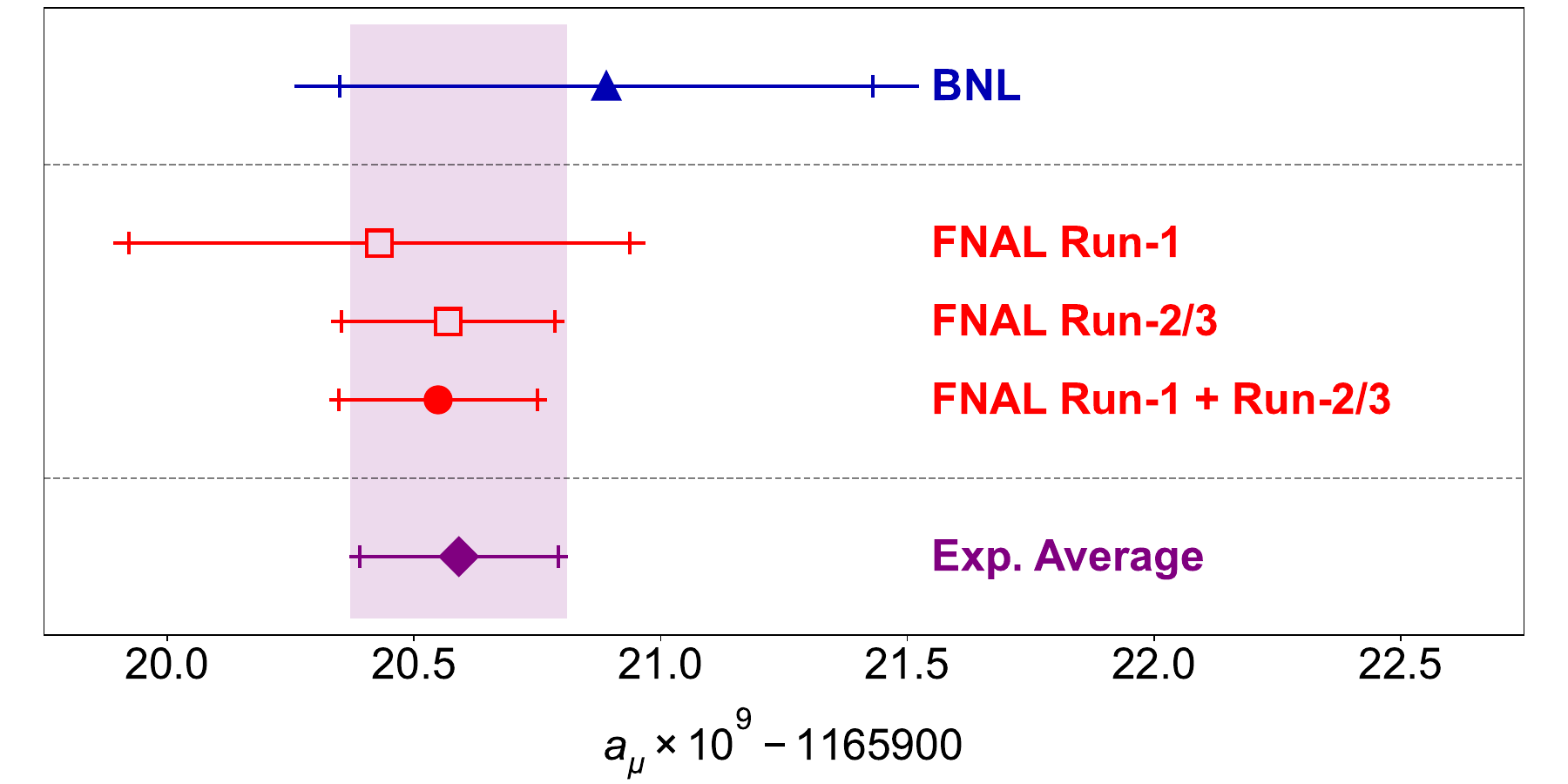} \\[-15pt]
      \begin{picture}(0,0)
        \put(12,25){(a)}
      \end{picture}
      \caption{(a) Plot of the Run-1 and Run-2/3 E989 results for $a_\mu$, 
        alongside the E821 final result, as well as an overall average.
        (b) FNAL E989 combined $a_\mu$ result to date, and the experimental
        world average (2023) are compared with three SM calculations.  See
        text for discussion.
        \label{fig:amu-avg}
               }
                           }
    \hspace*{\fill}
    \parbox{0.53\linewidth}{
      \includegraphics[width=\linewidth]{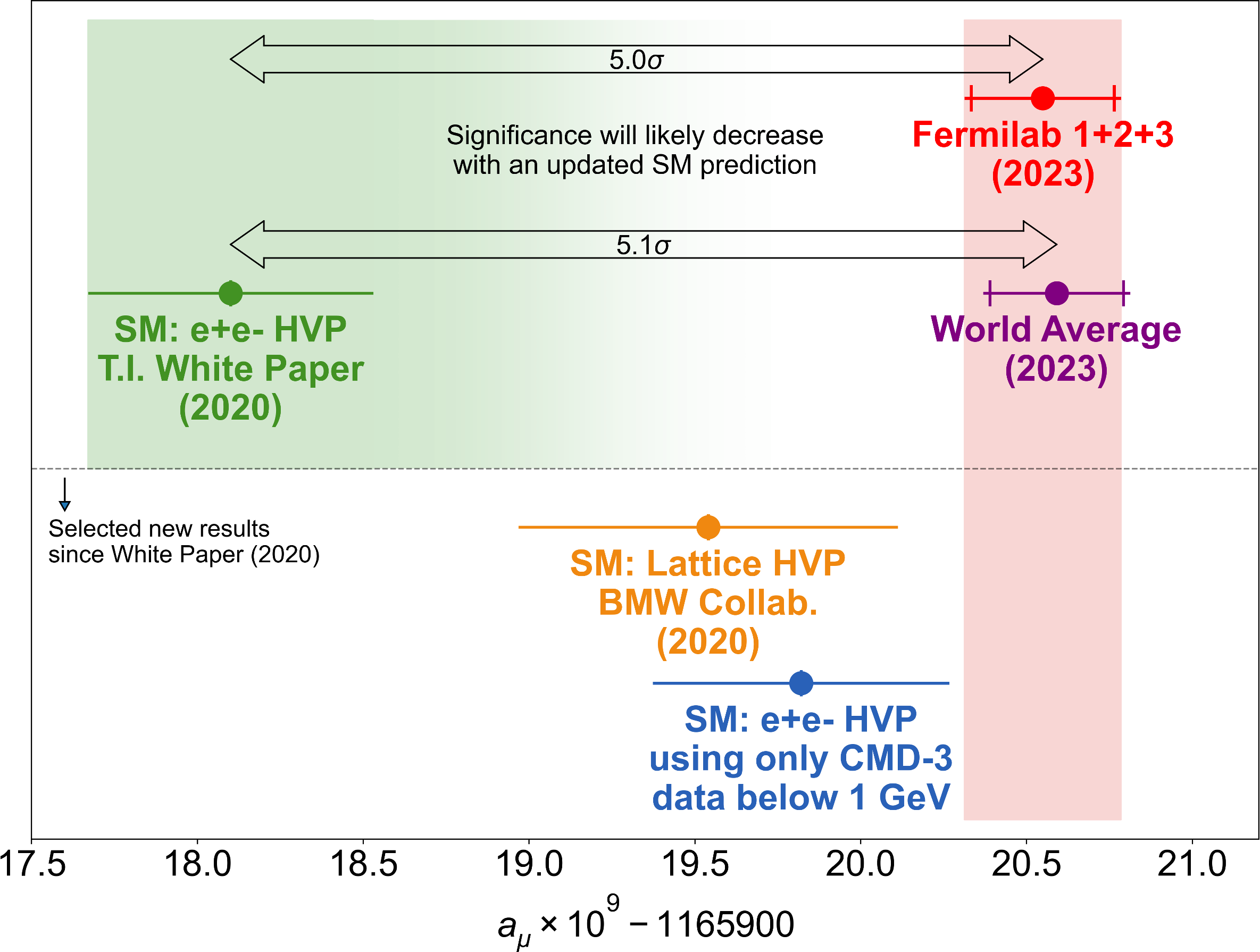} \\[-15pt]
      \begin{picture}(0,0)
        \put(12,30){(b)}
        \color{Peach}
        \put(178,88.5){\circle*{4}}
        \put(201,88.5){\line(-1,0){48}}
        \put(159.5,81.8){\tiny \textsf{\bfseries BMW (2024)}}
      \end{picture}
                           }
\end{figure}

The Run-1 unblinding took place in April 2021, and the unblinding for Runs 2
and 3 in July 2023.  The results are summarized in Figure~\ref{fig:amu-avg}.
Two observations readily stand out.  First, the E989 results are remarkably
internally consistent, and they are also consistent with the BNL E821 result
[Figure~\ref{fig:amu-avg}(a)].  On the other hand, the discrepancy relative to
the 2020 Muon $g-2$ TIWP result has only grown as new E989 results have become
available, and currently exceeds $5\sigma$.  Second, the experimental
uncertainties are lower than the SM uncertainties in the 2020 TIWP.  

However, the comparison with theoretical predictions is considerably muddied
by the 2021 Lattice QCD result by the BMW
collaboration\cite{Borsanyi:2020mff}, which predicted a value for $a_\mu$ less
than $2\sigma$ away from the experimental world average value.  The 2021 BMW
Lattice result is additionally reinforced by the 2023 experimental result for
$\sigma(e^+e^- \to \pi^+\pi^-)$\,\cite{CMD-3:2023alj} for $\sqrt{s} =0.32$ --
1\,GeV, which significantly differs from the world average
$\sigma_{\text{had}}(s)$ at comparable $\sqrt{s}$.  Finally, the most recent
update of their Lattice QCD calculation by the BMW
collaboration\,\cite{Boccaletti:2024guq}, marked as ``BMW (2024)'' in
Figure~\ref{fig:amu-avg}, reduces their 2021 uncertainty by 40\% and moves the
central value closer to experiment, to within $0.9\sigma$ of the current world
average for $a_\mu$.  Thus, there appears to be mounting evidence that the
$\sigma_{\text{had}}(s)$ data set underpinning the SM evaluation of $a_\mu$ by
the $g-2$ Theory Initiative needs to receive additional scrutiny.

Considerable experimental and theoretical effort will need to be expended in
order to clarify the current inconsistencies between the dispersion relation
based SM predictions of $a_\mu^{\text{HVP}}$ and the Lattice QCD calculations
of the same, resolve increasing inconsistencies apparent in the
$\sigma_{\text{had}}(s)$ data set, and to complete the E989 data analysis.
All of this work is ongoing.
In addition, interesting new experimental programs have been launched in order
to provide independent measurements of $a_\mu$ by different means and with
radically different systematics.

Since 2009, an international collaboration has been pursuing a different
measurement from the CERN/BNL/FNAL approach of storing muons at the ``magic''
momentum of 3.1\,GeV/$c$.  The JPARC Muon $g-2$ experiment starts with a
cooled muon beam and accelerates it into a tighter storage ring with no need
for electrostatic focusing.\cite{Abe:2019thb} The downside of this worthwhile
project is that it will not be able to match or exceed that of E989, which
limits it primarily to a consistency check.   It is notable, though, that the
JPARC experiment will also incorporate a measurement of the muon electric
dipole moment with competitive sensitivity.

The MUonE experiment currently preparing to take its Phase-1 beam time in
2025, approaches the problem in a different way: it aims to determine the
leading order term in $a_{\mu}^{\text{HVP}}$ (the dominant source of the SM
theoretical uncertainty) through precise measurement of muonic Bhabha
scattering, with precision sufficient to address meaningfully the above
inconsistencies.

\section{The MUonE experiment at CERN \label{sec:MUonE}}

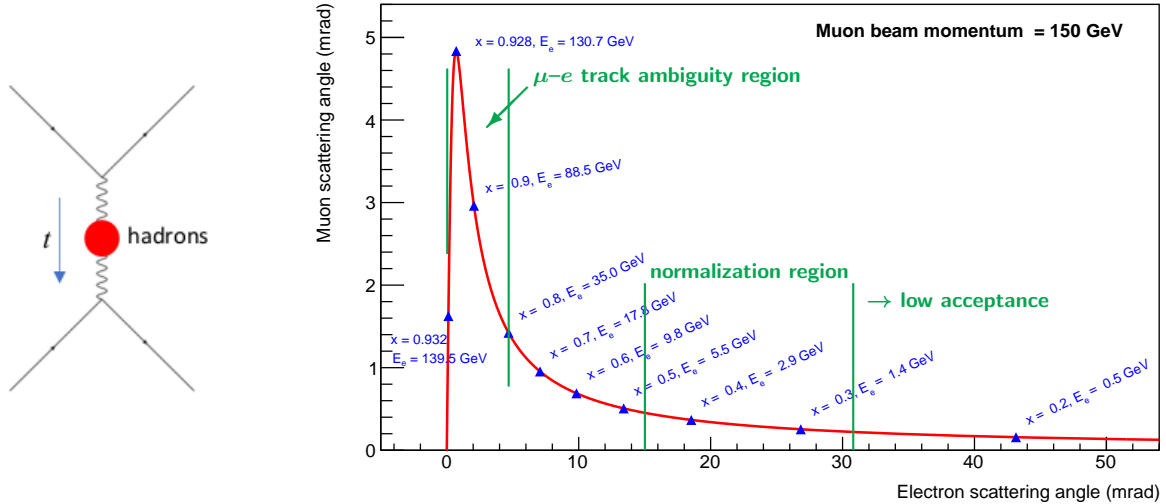
\begin{figure}
  \hspace*{\fill}
  \parbox{0.25\linewidth}{
    \includegraphics[width=0.7\linewidth]{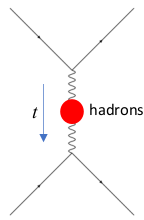}\hspace*{\fill}

    \vspace*{24pt}
                         }
  \parbox{0.7\linewidth}{
    \includegraphics[width=\linewidth]{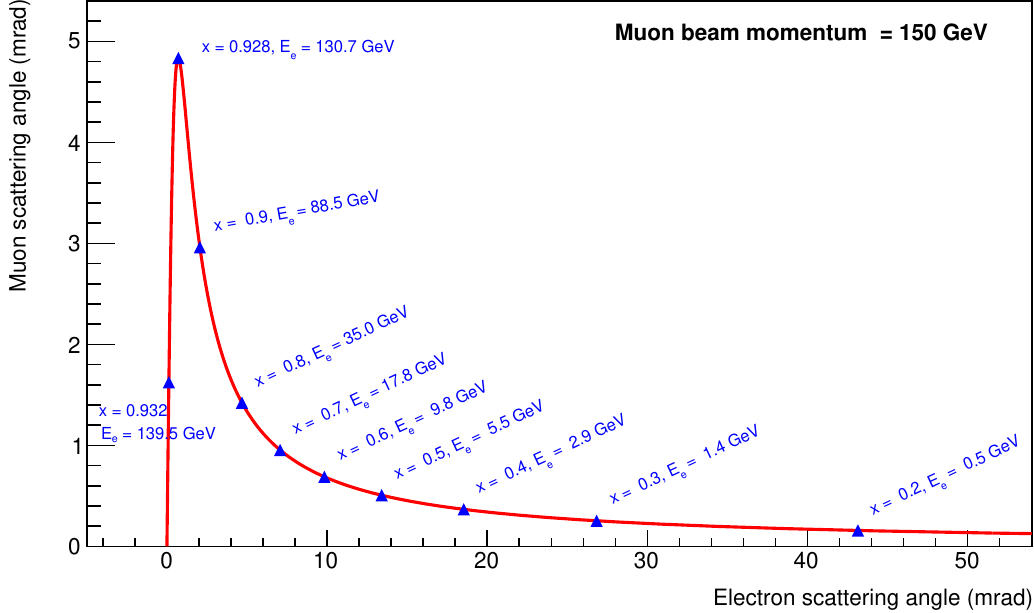} \\[-3pt]
    \hspace*{5pt}
    \begin{picture}(0,0)
      \sffamily\bfseries\thicklines\boldmath
      \put(43,175){\color{Green}\line(0,-1){70}}
      \put(66,175){\color{Green}\line(0,-1){120}}
      \put(75,170){\color{Green}\scriptsize
                           $\mu$--$e$ track ambiguity region}
      \put(73,168){\color{Green}\vector(-1,-1){15}}
      \put(117,94){\color{Green}\line(0,-1){63}}
      \put(195,94){\color{Green}\line(0,-1){63}}
      \put(119,96){\color{Green}\scriptsize
                           normalization region}
      \put(200,85){\color{Green}\scriptsize $\rightarrow$ low acceptance}
    \end{picture}      }
  \hspace*{\fill}

  \vspace*{-12pt}
 \caption{The leading order Feynman diagram of interest in MUonE (left).  The
   relation between the $\mu$ and the $e$ scattering angles for 150 GeV
   incident muon beam momentum (right). The blue triangles indicate the
   reference values of the variable $x$ and the electron energy, while green
   lines indicate the approximate boundaries of the three kinematic regions of
   interest in the experiment.} 
\label{fig:kin_corr}
\end{figure}

The MUonE collaboration, constituted in 2018, aims to complement the Muon
$g-2$ program by experimentally addressing the leading-order HVP
contribution $a_\mu^{\text{HLO}}$, the largest source of theoretical
uncertainty in $a_\mu$ \cite{MUonE:2019qlm}.  In light of the preceding
discussion and results shown in Fig.~\ref{fig:amu-avg}(b), that decision
appears prescient.
Using the theoretical relationship discussed in
Refs.~\cite{CarloniCalame:2015obs,deRafael:2020uif}, MUonE will evaluate the
hadronic contribution $\Delta\alpha_{\text{had}}$ to the running of $\alpha$,
the electromagnetic coupling, in a precise measurement of muon Bhabha
scattering on atomic electrons using a 160\,GeV muon beam in the M2 (North
Area) beamline at CERN.  MUonE's focus is on the leading order (LO) hadronic
contribution to $a_{\mu}$ via the relation~\cite{CarloniCalame:2015obs}
\begin{equation}
   a_{\mu}^{\rm HLO} = 
     \frac{\alpha}{\pi} \int_0^1 dx \, (1-x) \,  \Delta \alpha_{\rm had} \!
     \left[ t(x) \right], \quad\quad t(x)=\frac{x^2m_\mu^2}{x-1} < 0, 
\label{eq:amu_xalpha}  
\end{equation}
where $m_\mu$ is the muon mass and $\alpha$ is the fine-structure constant.
The hadronic contribution to the running of the effective electromagnetic
coupling, $\Delta\alpha_{\rm had}(t)$, evaluated at the space-like squared
four-momentum transfer $t(x)$, will be extracted from MUonE's precise
$\mu$-$e$ scattering measurements.  The $\mu$-$e$ two-body angular correlation
is shown in Fig.~\ref{fig:kin_corr}, along with the corresponding $x$ and
$E_e$ values.  The $\Delta\alpha_{\rm had}(t)$ term is strongest for
$\theta_{\mu,e} \lesssim 5$\,mrad, where the two cannot be unambiguously
separated on the basis of angle alone. 

While there are considerable theoretical advantages
over the dispersion approach through time-like processes: $e^+e^- \to
\text{hadrons}$, such as absence of resonances, this space-like scattering
approach presents formidable experimental challenges and requires highly
accurate simulations.

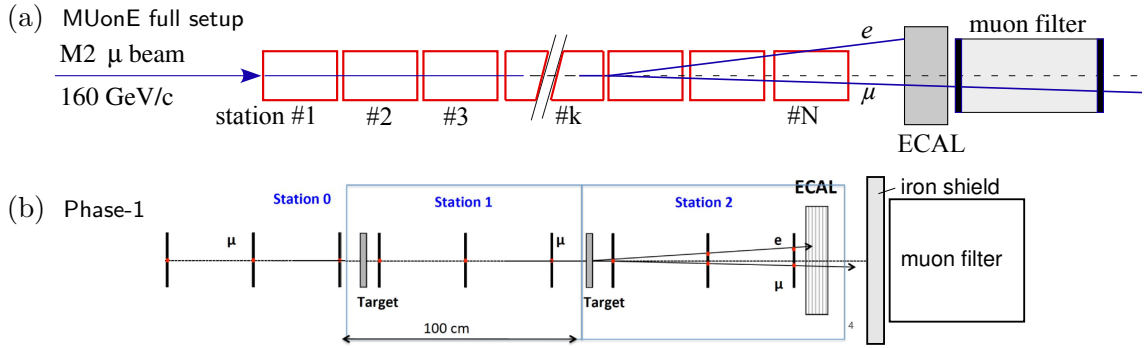
\begin{figure}[htb]
  \centering
  \parbox{0.9\linewidth}{
    \includegraphics[width=\linewidth]{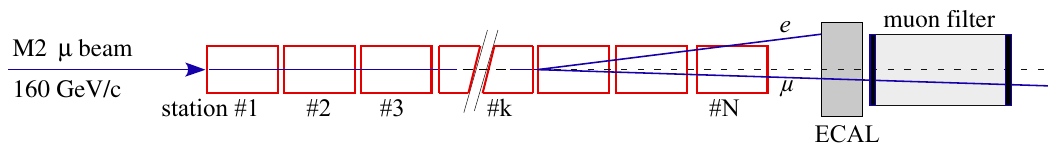} \\[3pt]
    \hspace*{0.1\linewidth}
    \parbox{0.64\linewidth}{
      \includegraphics[width=\linewidth]
                      {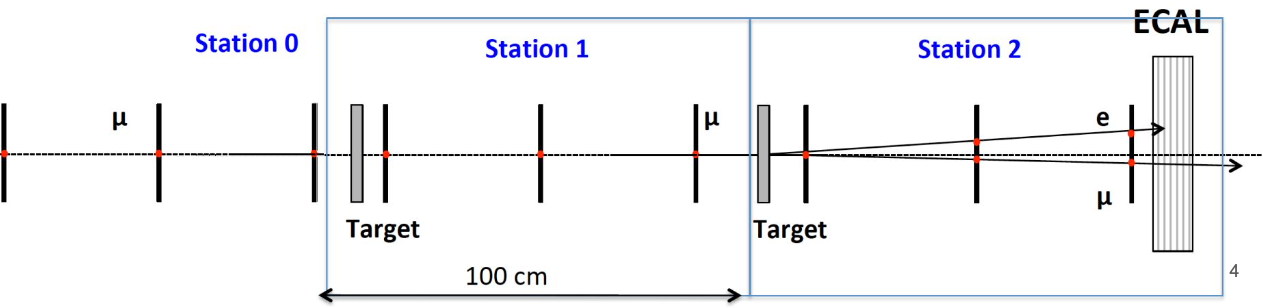}
                           }
    \hspace*{-8pt}
    \parbox{0.15\linewidth}{
      \includegraphics[width=\linewidth]{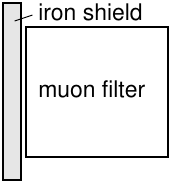}
                           }\\ 
    \begin{picture}(0,0)
      \put(-15,122){(a) \footnotesize\ {\sffamily MUonE full setup}}
      \put(-15,50){(b) \footnotesize\ {\sffamily Phase-1}}
    \end{picture}
                         }
    \caption{Layout sketches (not to scale) for: (a) the final MUonE apparatus
      with $N=40$ tracking stations, and (b) the Phase-1 measurements approved
      for 2025 in the M2 beamline.
      \label{fig:layout_fin_2025}
             }           
\end{figure}

\begin{figure}[hbt]  
   \parbox{0.74\linewidth}{
     \includegraphics[width=\linewidth]{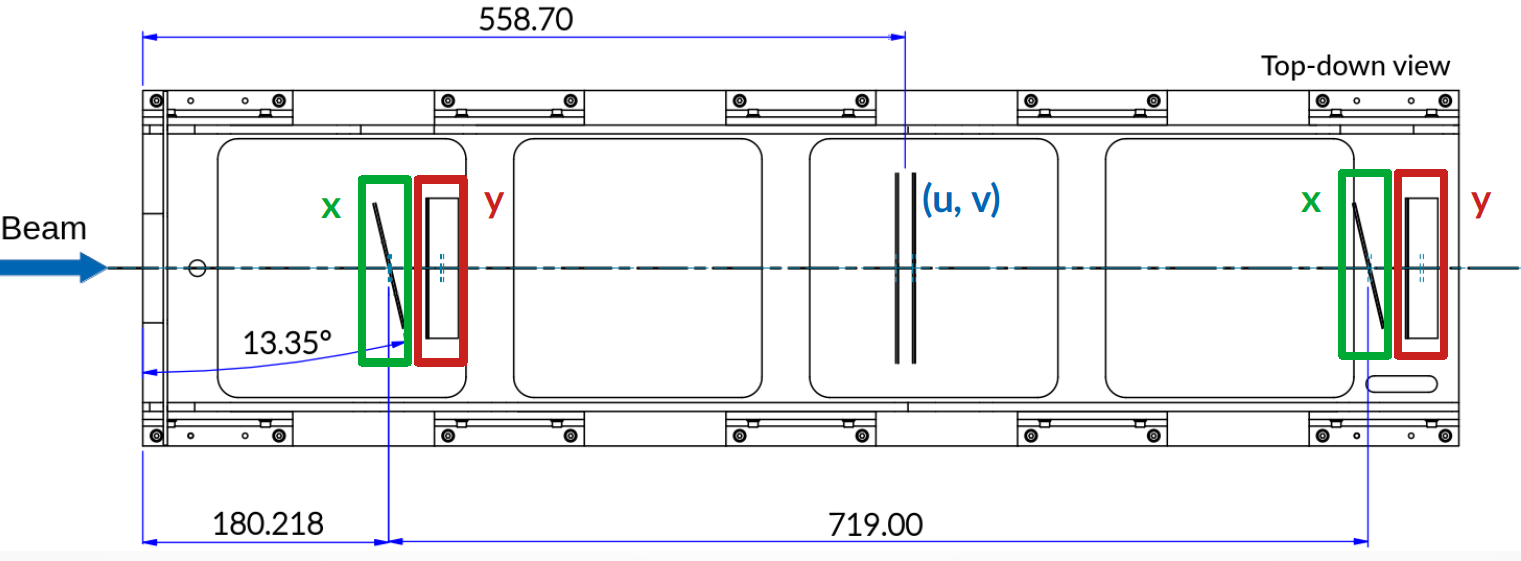}
                          }
   \hspace*{\fill}
   \parbox{0.24\linewidth}{
     \includegraphics[width=\linewidth]{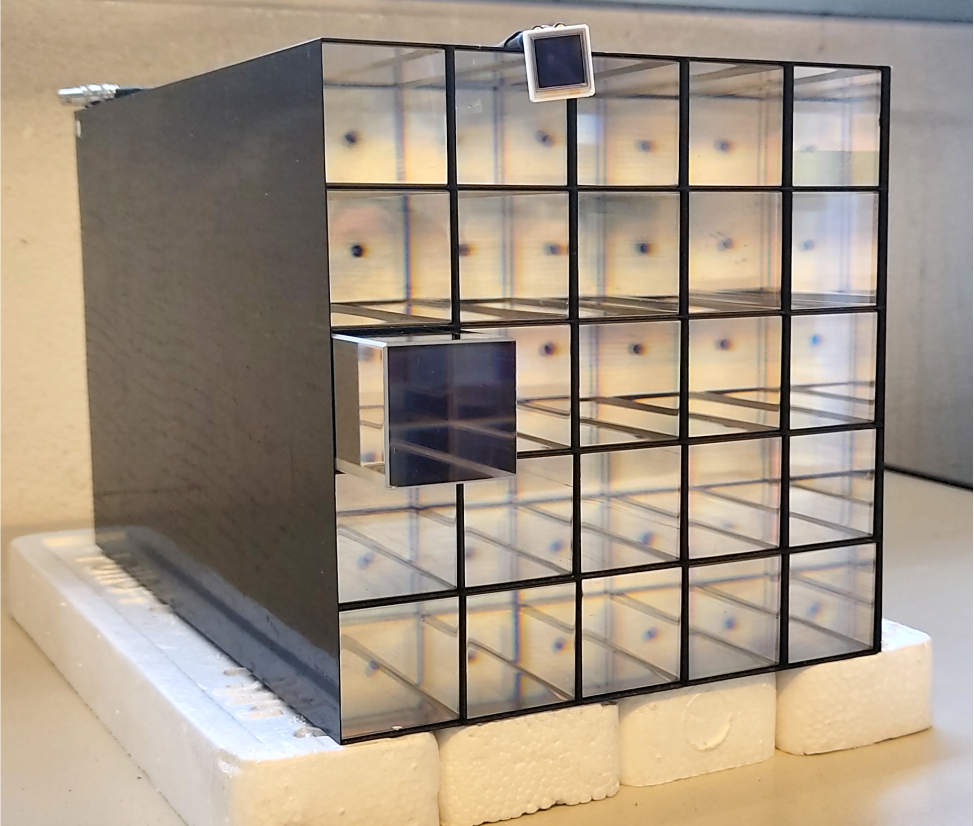}
                          }
   \caption{(a) An engineering drawing of a tracking station, showing how the
     $x$ and $y$ modules are tilted.  The graphite target is on the left, with
     its upstream face at the edge of the diagram. All dimensions are in mm.
     (b) Photograph of the 25 PWO crystals placed inside the carbon fiber
     alveolar matrix (downstream view). Resting on the top of the structure is
     a single $10 \times 10\,\text{mm}^2$ APD, used to detect the
     scintillation light.  } 
   \label{fig:TK_eng-ECAL}
\end{figure}

\begin{figure}[b!]
  \parbox{0.38\linewidth}{
    \includegraphics[width=0.9\linewidth]{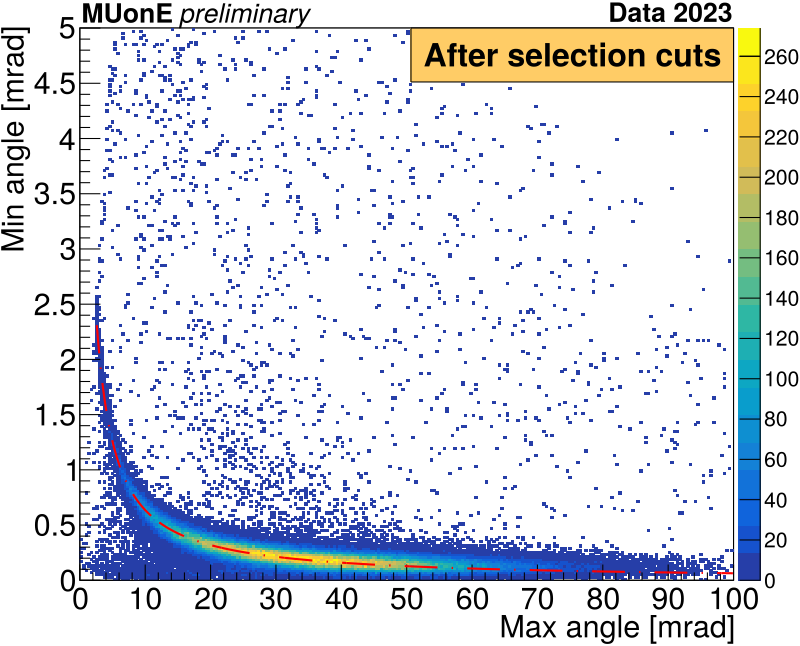} 
                         }
  \parbox{0.58\linewidth}{
    \includegraphics[width=\linewidth]{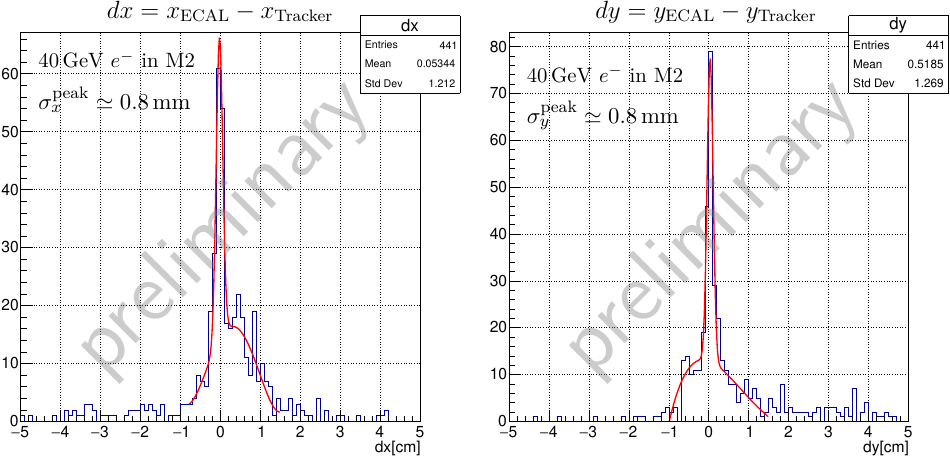}
                         } \\
  \begin{picture}(0,0)
    \put(120,90){(a)}
    \put(192,80){(b)}
    \put(328,80){(c)}
  \end{picture}
  \caption{Preliminary results from 2023 test beam data set.  (a)
    $\theta_{\text{min}}$ vs.\ $\theta_{\text{max}}$ distribution of elastic
    scattering candidates after the preliminary event selection.  The red
    dashed line represents the expected elasticity curve for a 160\,GeV muon
    beam.  (b) and (c) $x$ and $y$ spatial coordinate differences between the
    projected Tracker impact points on the ECAL front face, and the
    reconstructed electromagnetic shower centers, for synchronised
    Tracker-ECAL events (40\,GeV $e^-$ beam).  The near- Gaussian peaks, with
    $\sigma_{x,y}\simeq 0.8$\,mm, are consistent with Monte Carlo simulations
    of the same setup.}
    \label{fig:trac-ecal_spat}
\end{figure}
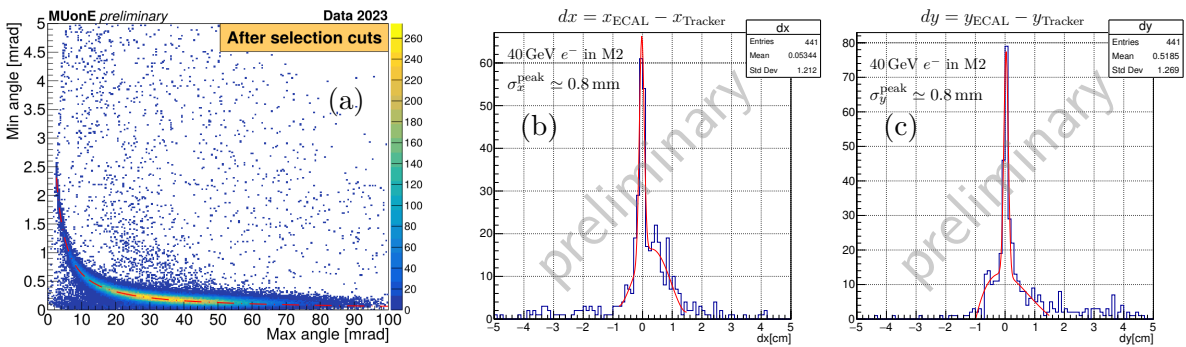

MUonE requires control of systematics at $\sim$10\,ppm level in order for the
subtraction of the leptonic, top-quark, and weak contributions to
$\Delta\alpha$ to be effective.  This calls for a light (Be) target segmented
into 40 slices, each followed by a $\sim$1\,m long tracking station; an
electromagnetic calorimeter (ECAL) and $\mu$ detector are at the end of
apparatus \cite{MUonE:2019qlm}.  MUonE aims to achieve $\sim$0.3\% statistical,
and $<$0.5\% overall relative uncertainty in $a_{\mu}^{\text{HLO}}$ in $\sim$2
years of data taking.  Layout of the full MUonE apparatus is sketched in
Fig.~\ref{fig:layout_fin_2025}, as well as the Phase-1 layout, to be used in
the 2025 beam period.
A closer look at the make-up of a tracker station and a look at the ECAL
$5\times 5$ PbWO$_4$ (PWO) calorimeter are found in
Figure~\ref{fig:TK_eng-ECAL}.  
At the heart of the experiment are the 2S tracking modules composed of two
single-sided Si microstrip detectors developed for CMS.\,\cite{CMS:2017lum}
The PWO single crystals were also originally used in the CMS
ECAL.\,\cite{CMS:2008xjf,CMS:2013lxn} For MUonE, the ECAL scintillation light
is read out by APDs, while the front-end electronics is based on the
multi-gain preamplifier (MGPA) chips \cite{Raymond:2005jm} developed for the
CMS ECAL.  Instrument goals include energy resolution $\sigma(E)/E \sim 1\%$
above 100 GeV and position resolution of $\lesssim 1$\,mm for the
reconstructed electron impact point.
MUonE has carried out a number of beam tests with individual components of the
current apparatus, as well as a complement of two or three tracking stations
and ECAL.  A few representative preliminary results of these measurements are
depicted in Figure~\ref{fig:trac-ecal_spat}.
As evinced in the figure, the key components of the MUonE apparatus have
met the basic requirements, and demonstrated the experimental proof of
principle. 

\section{\boldmath Conclusions: prospects for $a_\mu$ and hadronic vacuum
  polarization} 

In the coming year or two, the Fermilab E989 collaboration will unblind and
publish the results of analysis of the remaining three runs, Run-4 through 6,
or about 2/3 of the total data.  A significant improvement in precision is
expected over their Run-1/3 results, which have already exceeded the overall
goal for the systematic uncertainty.  Evidence appears to be increasing for a
real discrepancy between the Lattice QCD and dispersion-relation
determinations of $a_\mu^{\text{HLO}}$.  In the near future, strong ongoing
efforts on the theoretical side, as well as the new experiments: MUonE at CERN
and Muon $g-2$/EDM experiment at J-PARC, will provide much needed fresh
experimental evidence with radically different methods and systematics than
the previous experiments.

\section*{Acknowledgments}

This work has been supported by grants from the US National Science
Foundation, most recently PHY-2209484.

\end{document}